\documentstyle[12pt]{article}
\title{\bf Is magnetic flux quantized inside a solenoid? }
\author{F. Darabi \thanks{e-mail:f.darabi@azaruniv.edu} \\
{\small Department of Physics, Azarbaijan University of Tarbiat
Moallem, 53714-161 Tabriz, Iran .}}
\begin{document}
\maketitle
\begin{abstract}
In some textbooks on quantum mechanics, the description of flux
quantization in a superconductor ring based on the Aharonov-Bohm
effect may lead some readers to a (wrong) conclusion that flux
quantization occurs as well for a long solenoid with the same
quantization condition in which the charge of cooper pair $2e$ is
replaced by the charge of one electron $e$. It is shown how this
confusion arises and how can one avoid it.
\end{abstract}

\newpage

\section{Introduction}

Aharonov-Bohm effect is a quantum mechanical phenomena in which a
charged particle in a region free of magnetic field is affected by
the vector potential which produces that magnetic field \cite{1}.
One immediate result of this effect is to explain the magnetic flux
quantization in a superconductor ring. This is done based on the
uniqueness of the wave function after a $2\pi$ rotation around the
direction of magnetic field. One may usually think that such a
description is also applied for a magnetic field trapped inside a
long solenoid so that the corresponding flux is quantized as well as
superconductor case. However, it is easily shown that in case of
such quantization in a long solenoid applied in the Aharonov-Bohm
effect, we would not observe the shift of interference pattern in
this effect. In this short article, we try to understand the reasons
for flux quantization in the superconductor ring and flux
non-quantization in a long solenoid.

\section{Aharonov-Bohm effect}

The Schrodinger equation for a charged particle $e$ in the presence
of an electromagnetic field $(\textbf{A}, \phi)$ is given by
\begin{equation}
\frac{1}{2m}(-i\hbar \nabla -\frac{e}{c}\textbf{A})^2 \psi+e\phi
\psi =E \psi. \label{1}
\end{equation}
This equation is invariant under gauge transformation
\begin{equation}
\textbf{A} \rightarrow \textbf{A}+\nabla \lambda, \label{2}
\end{equation}
where $\lambda$ is an arbitrary function of space and time. This
invariance causes the wave function to acquire an extra phase, in
passing along a path through a region free of magnetic field,
according to
\begin{equation}
\phi=\frac{e}{\hbar c} \int_p \textbf{A}.\textbf{dx}. \label{3}
\end{equation}
The phase difference between the two cases where the charged
particle is moved along two different pathes with the same endpoints
encompassing the magnetic flux $\Phi$ is given by
\begin{equation}
\Delta \phi =\frac{e}{\hbar c} \Phi. \label{4}
\end{equation}
One may observe this phase difference by locating a long solenoid
having a variable magnetic flux between the two slits in the two
slits experiment. In fact, the relative phase difference of the wave
functions at a given point on the screen when the particle
(electron) passes the first or the second slit depends on the
magnetic flux encompassed by the long solenoid and any change in the
flux results in a shift in the interference pattern which is
observable as the Aharonov-Bohm effect.

\section{Flux quantization in a superconductor}

Consider a ring of superconductor containing a trapped magnetic flux
with a constant number of flux lines. The Schrodinger equation for a
unit charged particle, namely the cooper pair $2e$ is the same as
(\ref{1}) with $e$ replaced by $2e$. The wave function of this pair
after a $2\pi$ rotation around the direction of magnetic field
inside the ring is given by
\begin{equation}
\phi=\frac{2e}{\hbar c} \int_p
\textbf{A}.\textbf{dx}=\frac{2e}{\hbar c} \Phi. \label{5}
\end{equation}
The uniqueness of wave function then requires this phase to be an
integer multiplications of $2 \pi$ so that the magnetic flux is
quantized as follows
\begin{equation}
\Phi=\frac{2 \pi \hbar c}{2e}n, \:\:\: n=0, \pm1, \pm2,
....\label{6}
\end{equation}
In other words, the numbers of trapped flux lines can just be
integer multiplications of $\frac{2 \pi \hbar c}{2e}$. Flux
quantization in a superconductor was experimentally observed in 1961
by Deaver and Fairbank \cite{2}.

\section{Flux non-quantization in a long solenoid}

In some standard textbooks on quantum mechanics \cite{3} the subject
of flux quantization and Aharonov-Bohm effect are so vaguely
presented that one may mistakenly realizes the flux quantization is
a specific property of any magnetic flux such as the one inside a
long solenoid, whereas this is not the case and flux quantization
occurs just inside a superconductor ring. On the other hand, in some
other books the flux non-quantization in a long solenoid is
explicitly mentioned without a detailed explanation \cite{4}.

Consider a two slit experiment with a long solenoid, having a
magnetic flux $\Phi$, located between the two slits. At a given
point on the screen, the superposition of two wave functions of the
charged particle received by two slits $1, 2$ in the presence of the
long solenoid is given by
\begin{equation}
\psi=(\psi_1 e^{ie\Phi/\hbar c}+\psi_2)e^{{ie/\hbar c}\int_2
\textbf{A}.\textbf{dx}}. \label{7}
\end{equation}
The magnetic flux $\Phi$ is then responsible for a relative phase
difference between the two wave components received by two slits $1,
2$. This relative phase may shift the interference patter and this
is experimentally observed. It is immediately seen that if the
magnetic flux would be quantized according to
\begin{equation}
\Phi=\frac{2 \pi \hbar c}{e}n, \:\:\: n=0, \pm1, \pm2, ....\label{8}
\end{equation}
the above phase difference would be vanished and no shift in the
interference patter would be observed, a result which is in sharp
contrast with the observation. Therefore, we conclude that flux
quantization does not occur inside a long solenoid. However, a
question is immediately raised: Why is that the uniqueness of wave
function of the charged particle $e$ after a $2 \pi$ rotation around
the long solenoid does not lead to the flux quantization (\ref{8})?

\section{Simply and non-simply connectedness }

The wave function of the cooper pair inside a superconductor ring
should vanish on the interior walls of the ring and it simply means
that there is no cooper pair outside the ring, especially in the
interior hole encompassing the magnetic flux. Since there is no wave
function outside the ring then the closed path integral around the
magnetic flux is limited and confined to the interior region of the
ring where there are cooper pairs. This means the space is
non-simply connected so that one can not continuously contract the
closed integral to zero. Therefore, the phase acquired by the wave
function in this case is a topological one and the uniqueness
condition of wave function leads the magnetic flux to be quantized
according to (\ref{6}).

In the case of a long solenoid, however, there is no limitation or
confinement on the closed integral taken over a closed path around
the solenoid because there is no limitation on the presence of
electron in the region of magnetic flux inside the solenoid. This is
because, contrary to the superconductor ring there is no boundary
condition imposed on the wave function of an electron. In principle,
after a $2 \pi$ rotation of an electron around the long solenoid one
can continuously contract the resultant closed integral (total
phase) to zero with no difficulty, because the space in this case is
simply connected, namely the electron in principle can be present
every where in space. Therefore, the total phase is no longer a
topological phase and so one can not expect a quantization condition
(\ref{8}), rather we observe a continuous magnetic flux whose
continuous variation may lead to the Aharonov-Bohm effect in the two
slits experiment including a long solenoid containing magnetic flux.
It is worth noticing that although the magnetic flux in the
superconductor ring is quantized but the Aharonov-Bohm effect is
observed for a superconductor ring, as well. This is because, the
flux lines are integer multiplications of $\frac{\pi \hbar c}{e}$
and if we substitute it into the relative phase in (\ref{7}) the
phase shift occurs for the odd numbers of flux lines.

\section*{Acknowledgment}

The author would like to thank very much Prof. M. Berry for his
useful comments.


\begin{thebibliography}{99}
\bibitem{1}Y. Aharonov, and D. Bohm, Phys. Rev. 115, 485, 1959.
\bibitem{2}B. S. Deaver, and W.M. Fairbank, Phys. Rev. Letters. 7, 43, 1961.
\bibitem{3}S. Gasiorowicz, Quantum Physics, John Wiley and Sons, Inc,  1974;
J. J. Sakurai, Modern Quantum Mechanics, Addison-Wesley, Inc, 1985.
\bibitem{4}L. E. Ballentine , Quantum Mechanics, Prentice-Hall International , Inc, 1990;
M. Nakahara, Geometry, Topology and Physics, Adam Hilger, Inc,
1990.
\end{thebibliography}
\end{document}